\documentclass[iop, twocolumn]{aastex6}
\usepackage{amsmath}     
\usepackage{wasysym}           
\usepackage{graphicx}
\usepackage{amssymb}
\usepackage{epstopdf}
\usepackage{mathrsfs}
\usepackage{natbib}
\usepackage{color}
\usepackage{lipsum}

\begin{document}
\title{EXor outbursts from disk amplification of stellar magnetic cycles}
\author{Philip J. Armitage\altaffilmark{1,2}}

\altaffiltext{1}{JILA, University of Colorado and NIST, 440 UCB, Boulder, CO 80309-0440}
\altaffiltext{2}{Department of Astrophysical and Planetary Sciences, University of Colorado, Boulder}

\email{pja@jilau1.colorado.edu}

\begin{abstract}
EXor outbursts---moderate-amplitude disk accretion events observed in Class I and 
Class II protostellar sources---have time scales and amplitudes that are consistent with the 
viscous accumulation and release of gas in the inner disk near the dead zone boundary. 
We suggest that outbursts are indirectly triggered by stellar dynamo cycles, via 
poloidal magnetic flux that diffuses radially outward through the disk. Interior 
to the dead zone the strength of the net field modulates the efficiency of angular momentum 
transport by the magnetorotational instability. In the dead zone changes in the polarity 
of the net field may lead to stronger outbursts because of the dominant role of the Hall effect 
in this region of the disk. At the level of simple estimates we show that changes to kG-strength stellar 
fields could stimulate disk outbursts on 0.1~AU scales, though this optimistic conclusion depends 
upon the uncertain efficiency of net flux transport through the inner disk. The model 
predicts a close association between observational tracers of stellar magnetic activity 
and EXor events.
\end{abstract}

\keywords{accretion, accretion disks --- planets and satellites: formation --- protoplanetary disks --- instabilities} 

\section{Introduction}
EXor outbursts are repetitive moderate-amplitude accretion events observed in Class I and 
Class II protostellar sources \citep{herbig89}. The prototypical source, EX Lupi, is a $0.6 \ M_\odot$ 
M star with a moderate-mass disk and a quiescent luminosity of $0.7 \ L_\odot$. Outbursts in 
EXors have durations of the order of a year, recurrence time scales of a few years, and amplitudes 
that can be as large as $\Delta V \approx 5 \ {\rm mag}$ \citep{audard14}.

The origin of EXor outbursts is something of a mystery. They resemble 
anaemic versions of FU Orionis outbursts \citep{hartmann96}, which can be modeled 
as eruptive accretion disk phenomena localized to the inner AU of the system \citep{zhu09}. 
Indeed, with improving statistics on episodic accretion in pre-main-sequence stars it has 
been suggested that FUors and EXors may be limiting cases of a 
continuum of events that occur with a range of amplitudes in both Class I and Class II 
systems \citep{audard14}. Any observational kinship between FUors and EXors, however, 
sits uneasily with theoretical models for these phenomena. \cite{hartmann96} noted 
that a continuous range of outbursts would be consistent with thermal instability 
models  \citep{bell94}, but this conclusion probably does not 
carry forward to more recent models for FU Orionis outbursts that involve a 
limit cycle in which the inner disk alternates between a cold ``dead zone" \citep{gammie96} 
and a rapidly accreting hot state, whose onset is triggered by gravitational instability 
\citep{armitage01,martin11}. Time-dependent calculations of such a 
gravo-magneto limit cycle can broadly match observations \citep{zhu09b,bae14}.  
It is unlikely that the same model could be scaled down for EXors, whose 
lower mass disks and modest outbursts disfavor any role for gravitational 
instability. The same is true of an alternate class of models that link FU Orionis 
events to gravitational instability in the outer disk \citep{vorobyov05}. Models that 
invoke Roche lobe overflow from close-in planets have more potential to produce 
phenomenology that spans FUor and EXor-class events \citep{nayakshin12}, but 
are less studied.

We propose that EXor outbursts in the inner disk are externally triggered by changes 
in the strength and polarity of the stellar magnetic field. Young stars---including EXors \citep{hamaguchi12}---are 
frequently magnetically active \citep{bouvier07}, and cyclic accretion could occur 
as a consequence of either a time-variable torque on the inner disk \citep{clarke95,armitage95}, 
or from instability of a trapped disk whose inner edge is near corotation \citep{dangelo10}. 
\cite{dangelo12} developed an explicit model for EXors based on the latter idea.
In addition to 
this direct influence differential rotation between the star and the inner disk inflates 
and ultimately opens up a fraction of the stellar field lines \citep{lyndenbell94,agapitou00}. 
Outward diffusion of this stellar flux can modulate 
the net flux threading the inner disk and dead zone. Under ideal magnetohydrodynamic (MHD) conditions, a 
weak net flux acts to stimulate accretion occurring as a consequence of the magnetorotational instability 
\citep{hawley95}. Slightly further out, in the dead zone, the Hall effect  \citep{wardle99,balbus01} is 
the most important non-ideal MHD process \citep[for reviews see][]{armitage11,turner14}. Simulations 
show that the efficiency of angular momentum transport in the Hall-dominated regime 
differs dramatically depending upon whether the net field is aligned or anti-aligned to the 
rotation axis \citep{lesur14,bai14,simon15}. We therefore suggest that EXor outbursts occur 
due to the MHD response of the inner disk to the changing strength and polarity of a 
stellar-derived net magnetic field.

\section{Disk amplification of stellar magnetic activity cycles}
We assume that a net vertical magnetic field, derived from the stellar field, threads the inner 
part of the Keplerian accretion disk. At radii just outside the magnetosphere the 
disk is conductive enough to behave as an ideal MHD fluid \citep[this requires 
that the alkali metals are thermally ionized;][]{gammie96}, while further out in the dead zone non-ideal 
MHD effects are dominant. In both regimes the effect of the net vertical field on accretion 
is conventionally parameterized by  the mid-plane ratio of gas to magnetic pressure,
\begin{equation}
 \beta_z \equiv \frac{P_{\rm gas}}{B_z^2 / 8 \pi}.
\end{equation}
We make use of the results of shearing-box simulations in which the $B_z$ entering into the above 
equation is the {\em initial} field threading the local domain \citep[e.g.][]{simon15,salvesen16}, and 
hence define $\beta_z$ using the gas pressure in the disk and only the external magnetic field 
derived from the star. Under ideal MHD conditions it is well-established that a net field enhances 
the saturated accretion stress \citep{hawley95}. A fit to the results of \cite{salvesen16} gives the 
dependence as,
\begin{equation}
 \alpha = 1.1 \times 10^1 \beta_z^{-0.53}.
\end{equation}
Here $\alpha$ is the Shakura-Sunyaev parameter describing angular 
momentum transport \citep{shakura73}.
In the absence of any net field numerical simulations suggest that the magnetorotational 
instability \citep{balbus98} yields $\alpha \approx 0.01-0.02$ \citep{davis10,simon12}. A net 
field then enhances the stress for $\beta_z \lesssim 10^5$. Roughly the same threshold 
applies to the dead zone, notwithstanding the different physics at work there. For example, 
at 1~AU in a minimum mass Solar Nebula \citep{hayashi81} 
disk \cite{lesur14}, using simulations that included all three non-ideal MHD effects, found that 
$\alpha \approx 0.05$ for an aligned field with $\beta_z = 10^5$, 
while $\alpha \approx 4 \times 10^{-4}$ for an anti-aligned field of the same strength.

The mid-plane gas pressure can be estimated from steady-state viscous disk theory. In steady-state 
the surface density $\Sigma$ is related to the accretion rate $\dot{M}$ via $\nu \Sigma = \dot{M} / 3 \pi$, 
where the kinematic viscosity $\nu = \alpha c_s^2 / \Omega$. Here $c_s$ is the sound speed and $\Omega$ 
the angular velocity. The mid-plane density $\rho_0 = (1 / \sqrt{2 \pi}) (\Sigma / h)$, 
where the scale height $h = c_s / \Omega$. We then have,
\begin{equation}
 P_{\rm gas} = \rho_0 c_s^2 = \frac{1}{\sqrt{18 \pi^3}} \alpha^{-1} \dot{M} \Omega^2 c_s^{-1}.
\label{eq_disk_pressure}
\end{equation} 
For an initial estimate of where in the disk net fields may impact accretion 
we simply extrapolate the stellar dipole field $B_z = B_* (r/r_*)^{-3}$ to the 
disk outside the magnetosphere, and solve for the critical radius interior to which $\beta_z < 10^5$. 
Of the quantities entering the definition of $\beta_z$ the strongest radial dependence comes from the 
dipole variation of the external field strength, followed by $\Omega^2$ in equation~(\ref{eq_disk_pressure}). 
We therefore ignore the weak radial scaling of the sound speed, finding (for a Solar mass star),
\begin{eqnarray}
 r_{\rm crit} &\simeq& 0.095 
 \left( \frac{\beta_z}{10^5} \right)^{1/3} 
 \left( \frac{B_*}{\rm kG} \right)^{2/3}
 \left( \frac{r_*}{1.5 r_\odot} \right)^2 
 \left( \frac{\alpha}{10^{-2}} \right)^{1/3} \nonumber \\
 &\times& \left( \frac{\dot{M}}{10^{-8} \ M_\odot \ {\rm yr}^{-1}} \right)^{-1/3} 
 \left( \frac{c_s}{2 \ {\rm km s^{-1}}} \right)^{1/3} \ {\rm AU}.
\label{eq_rcrit} 
\end{eqnarray} 
Here we have used a sound speed $c_s = ( k_B T / \mu m_H)^{1/2} $ appropriate for $T \approx 10^3 \ {\rm K}$ 
(and mean molecular weight $\mu = 2.2$).

Interior to $r_{\rm crit}$ the response of the disk depends upon how the time scale for variations in the 
net flux compares to the viscous time scale. Rapid changes to the flux would modulate the local 
$\dot{M}$ at approximately constant $\Sigma$, while slow changes would lead to order unity 
variations in the surface density and mass of the inner disk (once it had attained a new steady-state). 
The amount of mass involved is $\Delta m \sim \pi r_{\rm crit}^2 \Sigma (r_{\rm crit}) \sim (1/3) \sqrt{GM_*} 
\alpha^{-1} \dot{M} r^{1/2} c_s^{-2}$. Adopting the same fiducial parameters as in equation~(\ref{eq_rcrit}) 
we find $\Delta m \sim 4 \times 10^{-6} \ M_\odot$. This is comparable to the amount of mass accreted 
during some EXor outbursts. \cite{audard10}, for example, estimated an accretion rate during outburst 
of $10^{-6} \ M_\odot {\rm yr}^{-1}$ for the EXor V1118 Orionis.

The above estimate suggests that changes to the stellar magnetic field, acting indirectly via its influence on 
the accretion stress in the inner disk, could potentially modulate the accretion rate at levels comparable 
to those required to explain EXor events. We now proceed to consider two additional complications. 
First, differential rotation between the star and its disk is likely to open up field lines, such that the relevant 
net field is an outwardly diffusing disk field rather than a direct extrapolation of the stellar dipole. Second, 
the naive estimate for $r_{\rm crit}$ above lies close to the inner edge of the dead zone \citep[nominally 0.1~AU 
in the original model of][]{gammie96}. A dynamically significant $\beta_z$ that extends into the dead zone 
opens up a stroner path to instability, because in the Hall-dominated inner disk not just the strength but also 
the {\em polarity} of the net field has a decisive impact on accretion \citep{lesur14,bai14,simon15}.

To estimate the disk field we assume that differential rotation opens up stellar field that 
threads the disk beyond the magnetospheric radius $r_m$. With a dipole approximation, 
the open magnetic flux that is available to diffuse through the disk is,
\begin{equation}
 \Phi_{\rm open} = 2 \pi B_* \frac{r_*^3}{r_m}.
\end{equation}
The detailed distribution of open flux in the inner disk will be time-dependent and 
complex, but is generically expected to have a flatter distribution than a simple 
extrapolation of the stellar dipole. The principle dependence is on the ratio of 
the effective velocities for the advection and diffusion of magnetic flux, $v_{\rm adv}$ 
and $v_{\rm diff}$ \citep{lubow94}. \cite{guilet14}  \citep[see also][]{okuzumi14} 
find approximate steady-state solutions for the disk field,
\begin{equation}
 B_z \propto r^{-n},
\end{equation}
with power-law slopes,
\begin{eqnarray}
 n & = & 0 \,\,\,\,\, v_{\rm adv} / v_{\rm diff} \ll 1 \nonumber \\
 n & = & 1 \,\,\,\,\, v_{\rm adv} / v_{\rm diff} \sim 1 \nonumber \\  
 n & = & 2 \,\,\,\,\, v_{\rm adv} / v_{\rm diff} \gg1.
\end{eqnarray} 
The appropriate values for the transport coefficients (and hence the ratio $v_{\rm adv} / v_{\rm diff}$) 
in protoplanetary disks are not fully determined, but the general theoretical preference 
is for $v_{\rm adv} / v_{\rm diff} \lesssim 1$ \citep{lubow94}. Leaving $n$ as a free parameter, we 
assume that after a reversal the open flux follows a power-law out to some distance $r$, with no 
flux at larger radii. The disk field at $r$ is then (for $n \neq 2$),
\begin{equation}
 B_{z,{\rm disk}} \simeq (2-n) B_* \frac{r_*^3}{r_m} r^{-2}.
\end{equation} 
Combining this with the expression for the radial dependence of the disk pressure 
(equation~\ref{eq_disk_pressure}) we find,
\begin{equation}
 \beta_z = \frac{8}{3 \sqrt{2 \pi}} \frac{GM_*}{(2-n)^2} \frac{1}{B_*^2 r_*^4} 
 \left( \frac{r_m}{r_*} \right)^2  \dot{M} \alpha^{-1} c_s^{-1} r,
\end{equation}
where $c_s$ and $\alpha$ are to be evaluated at radius $r$. Numerically, 
\begin{eqnarray}
 \beta_z = \frac{5.6 \times 10^4}{(2-n)^2} \left( \frac{M_*}{M_\odot} \right) 
 \left( \frac{B_*}{\rm kG} \right)^{-2} 
 \left( \frac{r_*}{1.5 r_\odot} \right)^{-4} \nonumber \\
 \times \left( \frac{r_m / r_*}{10} \right)^2 
 \left( \frac{\dot{M}}{10^{-8} M_\odot  {\rm yr}^{-1}} \right) 
 \left( \frac{\alpha}{10^{-2}} \right)^{-1} \nonumber \\
 \times \left( \frac{c_s}{2  {\rm km s}^{-1}} \right)^{-1}
 \left( \frac{r}{0.1 {\rm AU}} \right). 
\label{eq_beta_r} 
\end{eqnarray} 
This estimate is derived assuming a steady-state disk. To apply it to the dead zone, we need to know how 
the mid-plane pressure there differs from that in the adjacent thermally ionized region. 
One possibility is that $\alpha$ at the inner edge of the dead zone is very low, in which case gas could 
accumulate there producing high pressure conditions that would render plausible net disk 
fields permanently irrelevant. This would be the case if Ohmic diffusion were the only non-ideal 
effect at work \citep{gole16}. In the presence of the Hall effect, though, simulations suggest 
that the value of $\alpha$ given the favorable aligned orientation of the net field is {\em not} 
small, even in the nominal dead zone. Assuming frequent reversals of the net field's polarity, 
the appropriate $\alpha$ to use in the above estimate is the average of the 
aligned and anti-aligned Hall values, which is likely to be comparable to $10^{-2}$ \citep{lesur14}.

Adopting as the criterion for Hall effect physics to impact accretion at the inner edge of the dead zone 
that $\beta_z \lesssim 10^5$, equation~(\ref{eq_beta_r}) suggests that it is possible (but not guaranteed) 
for residual stellar fields diffusing through the disk to play a role. The factors that favor a 
role for stellar fields are lower disk accretion rates, strong  
stellar magnetic fields, and a larger fraction of the stellar flux available to diffuse 
through the disk. Exactly where the inner edge of the dead zone lies will also be 
critically important. The radius where the disk temperature first falls below the thermal ionization 
threshold of 800-1000~K 
\citep{gammie96,desch15} is dependent on the degree of viscous heating. A purely 
passive disk model \citep{chiang97} attains temperatures of $10^3 \ {\rm K}$ only 
well inside 0.1~AU. Models that include viscous heating (typically following an $\alpha$-prescription 
formalism, and assuming standard dust opacities) push the critical radius out to 
$r \simeq 0.3 \ {\rm AU}$ \citep{bell97}. Intermediate values would result if dead zone heating is 
strongly concentrated toward low optical depths near the disk surface.

\begin{figure}
\begin{center}
\includegraphics[width=0.45\textwidth,angle=0]{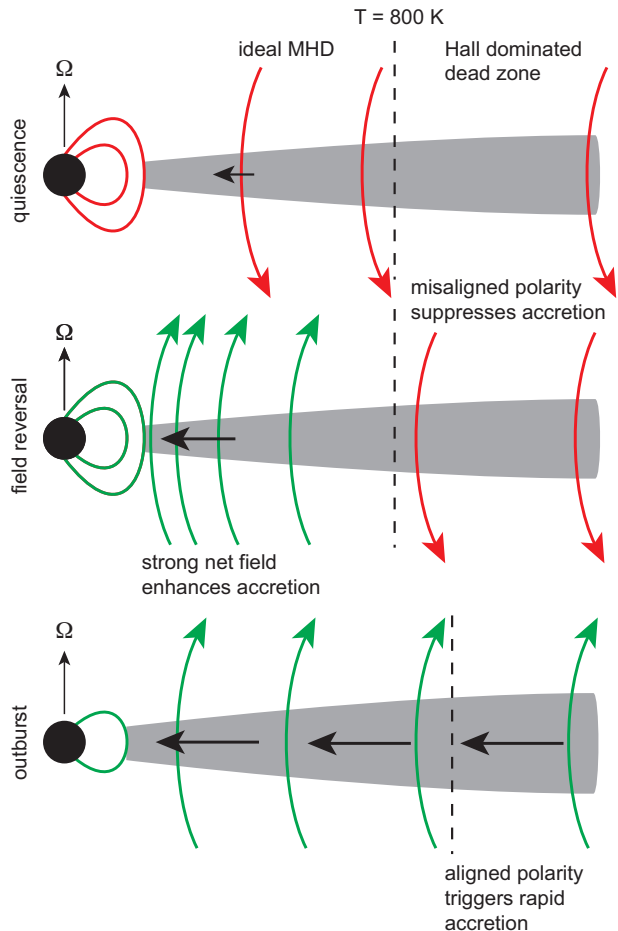}
\end{center}
\caption{An illustration of how stellar magnetic cycles may propagate through the inner disk into the dead zone 
and lead to accretion outbursts. The stellar dipole field dominates the dynamics within a magnetospheric radius 
$r_m \geq r_*$. Beyond the magnetosphere, differential rotation between the disk and stellar surface 
opens up a fraction of the stellar field lines, liberating vertical flux that diffuses outward through the 
disk at a rate much faster than viscous diffusion. This affects the accretion rate via two effects. 
In the region of the disk where ideal MHD is a 
good approximation, the accretion stress increases at points in the cycle where the {\em strength} 
of the net field is locally high. In the dead zone, the Hall effect strongly suppresses accretion when the 
{\em polarity} of the field is anti-aligned to the disk rotation. Subsequent diffusion of 
aligned field into the dead zone results in efficient angular momentum transport, heating, and a burst 
of accretion.}
\label{fig1}
\end{figure}

Figure~\ref{fig1} illustrates how stellar dynamo cycles could trigger accretion outbursts via 
the polarity-dependence of Hall MHD. Low-rate accretion occurs when the net disk field threading 
the inner edge of the dead zone, derived from the stellar field, is anti-aligned to the disk rotation 
$({\bf \Omega} \cdot {\bf B} < 0)$. High-rate accretion is triggered when the stellar magnetic 
field flips, and aligned net field $({\bf \Omega} \cdot {\bf B} > 0)$ diffuses outward into the dead 
zone region. As noted above the reversal in net field polarity could plausibly alter the efficiency 
of angular momentum transport by an order of magnitude, or more. We caution, however, 
that the impact of the Hall effect on disk accretion depends on details of the disk chemistry, 
and simulations relevant to the inner edge of the dead zone at radii interior to 1~AU have 
not been reported.

In addition to the field being strong enough, two additional conditions need to be met for 
stellar magnetic cycles to trigger dead zone EXor outbursts. First, changes in the stellar field 
need to be communicated to the dead zone 
region on a time scale of, at most, a few years, to match observational constraints on outburst 
recurrence times \citep{audard14}. Second, the viscous time scale in the outburst state 
needs to match observations.

To estimate the time scale for communication of stellar field changes to the dead zone we 
apply the transport theory developed by \cite{lubow94}, which is based on assuming that 
a purely poloidal force-free field threads a thin turbulent disk. In this limit diffusion of the 
net field occurs on a time scale that is shorter by a factor of $\sim (h/r)$ than radial 
advection by the accretion flow (implying, in this scenario, that net field liberated 
from the magnetosphere escapes radially outward). This results in an estimate,
\begin{eqnarray} 
 t_{\rm diff} & \sim & \frac{1}{\alpha \Omega} \left( \frac{h}{r} \right)^{-1} \nonumber \\
 & \sim & 2.4 \left( \frac{\alpha}{0.1} \right)^{-1} 
 \left( \frac{c_s}{2 {\rm km s}^{-1}} \right)^{-1} 
 \left( \frac{r}{0.1 {\rm AU}} \right) \ {\rm yr}.
\end{eqnarray} 
We adopt a higher value of $\alpha$ here because the field is diffusing through the inner 
disk, described by ideal MHD, where the net field enhances transport independent 
of field polarity \citep{hawley95}. 

This estimate is crude and it ignores important physical effects. The complex and time-dependent 
interaction between the stellar magnetosphere and the disk \citep{romanova08,blinova16} will 
affect net flux transport in the inner disk. Both MHD and force-free simulations show that 
open field is rapidly dispersed from the vicinity of the magnetosphere \citep{romanova09,parfrey16}. 
Outward transport at a rate substantially faster than that implied by the \cite{lubow94} model 
would reduce the field strength threading the dead zone, and limit the influence of the Hall effect 
on accretion. Working in the opposite direction, the effect of the vertical disk structure on the 
transport co-efficients can enhance (inward) advection with the mean flow \citep{lovelace09}. 
Finally the discontinuity in disk properties 
at the location of the dead zone may also impact transport, perhaps particularly in the case 
where the inner part of the dead zone is largely laminar. None of these effects are easy 
to quantify. We note, however, that an 
outburst with an accreted mass of $10^{-6} \ M_\odot$ requires a surface density change 
of only $\Delta \Sigma \sim 300 \ {\rm g \ cm}^{-2}$ over 0.1~AU scales. It seems 
possible that enough field could diffuse to 0.1~AU on $\sim$yr time scales to cause 
such a perturbation.

Potentially the hardest constraint to match is on the expected time scale of the events. If we assume 
that the $\sim 5$ magnitude increase in brightness during an EXor event points to roughly 
a 2 orders of magnitude increase in accretion rate, the disk sound speed  
at the onset of outburst might rise by a modest factor (very approximately we expect 
$c_s \propto \dot{M}^{-1/8}$). The nominal 
viscous time $t_\nu = r^2 / \nu$ is then,
\begin{eqnarray}
 t_\nu & \sim & \sqrt{GM_*} \alpha^{-1} c_s^{-2} r^{1/2} \nonumber \\
 & \sim & 28 \left( \frac{\alpha}{0.1} \right)^{-1} 
 \left( \frac{c_s}{4 {\rm km s}^{-1}} \right)^{-2} 
 \left( \frac{r}{0.1 {\rm AU}} \right)^{1/2} \ {\rm yr}.
\end{eqnarray}   
In practice the evolution of disk structures containing sharp initial gradients occurs on a 
small fraction of the nominal viscous time \citep[see e.g. the Green's function 
solution for a narrow ring;][]{pringle81}, so the above estimate is likely marginally consistent with 
EXor durations of the order of a year. Generically, if both FUors and EXors originate 
from disk instabilities associated with the dead zone we expect FUors to be longer 
lived events, consistent with observations, as the inner edge of the dead zone is further out at higher accretion 
rates.

\section{Discussion}
We have argued that EXor outbursts could be triggered by stellar dynamo 
cycles, via a mechanism that involves diffusion of stellar magnetic flux through the inner disk 
into the Hall-dominated dead zone. There are obvious uncertainties. 
In the region of the disk governed by ideal MHD we have ignored the possibility that net 
flux threading the inner disk launches a disk wind \citep{bai13}, 
and neglected the known complexities of disk-magnetosphere interactions \citep{romanova09,parfrey16}. 
In the dead zone we have motivated our arguments using simulations that are strictly 
appropriate only at larger radii, and although the physics underlying the Hall-mediated bimodality 
in accretion stress is well-established \citep{wardle99,lesur14,bai14,simon15} our conclusions 
are quite dependent on the location of the inner edge of the dead zone. More broadly, of course, 
it is not known whether EXor hosts exhibit the requisite dynamo activity.

Observationally, an association between EXors and stellar magnetic activity is an 
obvious possibility given the time scales, recurrences, and approximately periodic 
nature of the outbursts. Irrespective of the details, the model predicts that EXor events 
ought to correlate with indicators of 
stellar magnetic activity (with a lag reflecting the propagation time scale of 
magnetic field through the disk), and might be preferentially seen toward those 
stars with the strongest magnetic activity.

In the broader context of eruptive protostellar phenomena, some recent observational studies advocate a  
scenario in which FUors and EXors are members of a continuum of outbursting 
systems \citep{audard14}. Such a scenario appears consistent with a theoretical model in which FUors and EXors  
both arise from modulation of inner disk accretion related to instabilities that occur near the inner edge 
of the dead zone at 0.1-1~AU scales. What may distinguish EXors and FUors is the nature of the 
trigger, with EXors being externally triggered by changes in the 
stellar magnetic field, whereas FUors are internally triggered by the onset of 
gravitational instability \citep{armitage01}. 

\acknowledgments
I thank the referee for an insightful report, and acknowledge support from NASA through 
grants NNX13AI58G and NNX16AB42G, and from the NSF 
through grant AST 1313021.

\bibliographystyle{aasjournal}

\end{document}